\documentclass[%
reprint,
superscriptaddress,
amsmath,amssymb,
aps,
]{revtex4-1}
\usepackage{color}
\usepackage{graphicx}
\usepackage{dcolumn}
\usepackage{bm}
\usepackage{textcomp}

\usepackage{braket}

\usepackage{overpic}
\usepackage{comment}

\begin {document}

\title{Observation-Time-Induced Crossover in Driven Anomalous Transport}

\author{Masahiro Shirataki}
\affiliation{%
  Department of Physics and Astronomy, Tokyo University of Science, Noda, Chiba 278-8510, Japan
}%

\author{Takuma Akimoto}
\email{takuma@rs.tus.ac.jp}
\affiliation{%
  Department of Physics and Astronomy, Tokyo University of Science, Noda, Chiba 278-8510, Japan
}%



\date{\today}

\begin{abstract}
We investigate how a weak constant force becomes detectable through fluctuations in anomalous transport in strongly heterogeneous media. Rather than focusing on the mean drift, we show that the key signature of the force appears in the variance of the particle displacement.
As representative models, we study a driven continuous-time random walk (CTRW) with nearest-neighbor jumps and a driven quenched trap model (QTM) with a power-law waiting-time tail. 
By analysing the force dependence of the displacement variance, we quantify how fluctuations respond to weak driving.
We find that for $\alpha<2$, the response exhibits an observation-time-induced crossover: at fixed bias, the variance initially follows its unbiased scaling and only at later times crosses over to a force-dominated nonequilibrium regime. Equivalently, at fixed observation time $t$, there exists a threshold bias $\varepsilon_c(t)$ separating an apparently equilibrium-like regime from a detectable nonequilibrium response. This threshold decreases with increasing $t$, implying that weaker forces become observable over longer measurement windows.
Quenched disorder further lowers the detection threshold compared with CTRW, and the crossover reflects a competition between the finite observation time and the intrinsic relaxation time of the driven heterogeneous system.
\end{abstract}

\maketitle


\section{Introduction}

Transport in disordered systems is commonly characterized in the long-time limit, where steady-state response and asymptotic scaling laws are well defined. Under such conditions, transport coefficients and response functions are typically assumed to be independent of the observation protocol, an assumption that underlies linear-response theory and many standard approaches to nonequilibrium transport \cite{Kubo1957, RKubo_1966, Green1954}.

In practice, however, measurements are always performed over finite observation times, and the extracted transport coefficients may depend on the measurement window \cite{Mantegna1994,He2008,Weigel2011,Miyaguchi2011a,Miyaguchi2013,Akimoto2016a}. For instance, in water diffusion through porous media, the apparent diffusion coefficient varies with diffusion time, reflecting confinement and surface effects \cite{MitraSenSchwartz1993PhysRevB,HurlimannLatourSotak1994MRI}. These observations indicate that transport coefficients measured over finite time windows may differ substantially from those in the long-time regime. Here, “long-time” refers to the asymptotic scaling regime corresponding to the limit $t \to \infty$, whereas “finite observation time” refers to measurements performed at finite $t$, where the asymptotic scaling may not yet be established.

Such finite-time effects become even more pronounced in disordered and glassy systems exhibiting anomalous transport, where relaxation is slow and convergence toward asymptotic scaling is strongly delayed \cite{Berthier2011,bouchaud90,bouchaud:jpa-00246652,10.1063/1.1559676}. Continuous-time random walks (CTRW) and quenched trap models (QTM) provide paradigmatic descriptions of these systems, capturing broad waiting-time distributions and strong dynamical heterogeneity \cite{Shlesinger1982,bouchaud90,metzler00}. In these models, rare and long trapping events dominate the dynamics, leading to slow relaxation and long transient regimes before asymptotic transport behavior is established \cite{Wang2019,wang2020large,Akimoto2016a,Akimoto2018,Sakai2023L,Sakai2023}.

When an external driving force is applied, transport in these systems has been extensively studied in terms of steady currents and nonlinear response \cite{Scher1975,habdas2004forced,Gazuz2009,PhysRevE.67.065105,Gradenigo2016,Hou2018,Akimoto2018b,Akimoto2020trace,Wang2020,PhysRevLett.133.037101,shafir2022case,Sakai2023}. Most existing studies focus on the long-time transport regime, implicitly assuming that observation times are sufficiently long for asymptotic behavior to be well established. The role of finite observation time, however, has received far less attention, despite the fact that driven transport can remain far from its asymptotic form over experimentally relevant time scales. In particular, it remains unclear how weak driving becomes detectable in finite-time transport observables when the system exhibits strong dynamical heterogeneity.

In this work, we investigate how a weak constant bias becomes detectable through fluctuations in anomalous transport. Focusing on the displacement variance in driven CTRW and QTM systems with power-law waiting-time distributions, we show that finite observation time fundamentally alters the transport response. Specifically, the variance exhibits an observation-time-induced crossover: for sufficiently weak bias the fluctuations remain indistinguishable from the unbiased dynamics, whereas above a threshold bias the variance crosses over to a bias-induced scaling. This threshold bias decreases with the observation time, implying that weaker driving becomes detectable over longer observation windows. Our results establish observation-time-induced crossover as a robust feature of driven anomalous transport and provide a quantitative framework for diagnosing the finite-time detectability of weak driving in heterogeneous systems.
This crossover originates from the competition between the observation time and a bias-dependent relaxation timescale of the transport response.


\section{Models and Renewal-Theory Framework}

\subsection{Driven CTRW and QTM}

We investigate how quenched disorder affects the response to an external bias by comparing the driven CTRW and the driven QTM. These models serve as fundamental descriptions of subdiffusive transport in heterogeneous systems.
We consider an infinite one-dimensional lattice with unit spacing. A particle performs nearest-neighbor jumps, waiting for a random time drawn from a prescribed waiting-time distribution before each jump.
The stochastic dynamics can be formulated within a renewal framework, where the number of jumps up to time $t$ determines the transport statistics.

In the CTRW, at each jump the particle moves from site $x$ to \(x+1\) with probability \(p\) and to \(x-1\) with probability \(q=1-p\).
The waiting times $\tau$ between successive jumps are independent and identically distributed (IID) random variables drawn from a power-law probability density function (PDF),
\begin{align}
    \psi(\tau)=\frac{\alpha}{\tau_0}\left(\frac{\tau}{\tau_0}\right)^{-1-\alpha}\label{eq:wt_dist_pow}
\end{align}
where \(\tau_0\) is a characteristic time, set to 1 in the numerical simulations.
The exponent $\alpha$ controls the degree of dynamical heterogeneity: smaller values of $\alpha$ correspond to broader waiting-time distributions and stronger subdiffusive behavior. For $0<\alpha<1$, the mean waiting time diverges, whereas for $1<\alpha<2$, it remains finite but the variance diverges. 
In the present work, we consider the ordinary renewal process, where the dynamics starts at $t=0$ and the first waiting time is drawn from the same distribution as subsequent ones.

The QTM describes diffusion in a quenched random energy landscape, where the trap depth at each lattice site remains fixed on the timescale of the particle dynamics. Each site $n$ is characterized by an energy depth $E_n$, which determines the mean waiting time via the Arrhenius law,
\begin{align}
    \tau_n = \tau_0 \exp\!\left(\frac{E_n}{k_{\mathrm{B}} T}\right),
\end{align}
where $\tau_0$ is a characteristic timescale of the dynamics. In the numerical simulations, we set $\tau_0 = 1$.
The trap depths are independent random variables drawn from an exponential PDF, 
\begin{align}
    \rho( E)=\frac{1}{k_\mathrm{B}T_g}\exp\left(-\frac{ E}{k_\mathrm{B}T_g}\right),
\end{align}
where \(T_g\) denotes the glass temperature. Combining these assumptions yields a power-law waiting-time PDF over all sites,
\begin{align}
  \psi(\tau)=\frac{\alpha}{\tau_0}\left(\frac{\tau}{\tau_0}\right)^{-1-\alpha}\quad(\alpha\equiv T/T_g),
\end{align}
where \(\alpha \equiv T/T_g\). 
Unlike the CTRW, where waiting times are independently drawn at each jump, the QTM features quenched disorder: the waiting time is fixed by the site and revisited upon return. 
In the QTM, we also consider a non-equilibrium initial condition, where the particle is placed at a randomly selected site at $t=0$ without bias toward deep traps. This differs from the stationary state of the QTM, where the occupation probability is weighted by the trapping times, leading to aging effects.

Under an external bias, a particle at position $x$ waits for a time determined by the local trap depth and then jumps to $x+1$ with probability $p$ or to $x-1$ with probability $q=1-p$. 
We  introduce the parameter \(\varepsilon \equiv p - q\) to 
quantify the strength of the bias. In the weak-force limit, $\varepsilon$ is related to the external force $F$ through
\begin{align}
    \varepsilon=\frac{F}{2k_\mathrm{B}T}\quad(F\rightarrow 0),
\end{align}
which follows from the detailed-balance condition \cite{bouchaud90}.
In both the CTRW and QTM considered here, the jump probabilities $p$ and $q$ are independent of the lattice site. This contrasts with the Sinai model \cite{sinai1982limit}, where the bias is site dependent. The present setting is nevertheless relevant to transport in comb-like structures \cite{arkhincheev1991anomalous} and captures diffusion in a random energy landscape under weak constant driving.

\subsection{Displacement moments in terms of the jump number}

We first formulate the transport statistics in terms of the number of jumps up to time $t$. For the CTRW, the waiting times are IID, and the process is a renewal process. In the QTM, the waiting times are quenched and therefore not independent upon revisiting sites; however, the particle position can still be written as a sum over the number of jumps, allowing the same decomposition in terms of $N_t$.
Let $N_t$ be the number of jumps up to time $t$. Then
\begin{align}
  x(t)=\sum_{i=1}^{N_t}\delta x_i .
\end{align}
Since the step increments $\{\delta x_i\}$ are independent of of the waiting times and thus independent of the jump number $N_t$, Wald's identity yields
\begin{align}
  \Braket{x(t)}=\Braket{N_t}\Braket{\delta x},
\end{align}
and
\begin{align}
  \Braket{x(t)^2}
  =\Braket{N_t} \mathrm{Var} (\delta x )
  +\Braket{N_t^2}\Braket{\delta x}^2 .
\end{align}
In the presence of an external bias $\varepsilon$, we obtain
\begin{align}
  \Braket{x(t)}_{\varepsilon}
  &= \varepsilon \Braket{N_t}, 
  \label{eq:meandrift}\\
  \mathrm{Var}\!\left[x(t)\right]_{\varepsilon}
  &=(1-\varepsilon^2) \Braket{N_t}
  +\mathrm{Var}(N_t)\varepsilon^2 .
\end{align}
Rearranging terms gives
\begin{align}
  \mathrm{Var}\!\left[x(t)\right]_{\varepsilon}
  =\Braket{N_t}+\left(\mathrm{Var}(N_t)-\Braket{N_t}\right)\varepsilon^2 .
  \label{eq:VarXt_Nt}
\end{align}
Equation~\eqref{eq:VarXt_Nt} holds for continuous-time random-walk-type models in which the step increments are independent of the jump number $N_t$, and therefore applies to both the CTRW and the QTM considered here.

\section{Asymptotic expansion of the jump-number statistics}

\subsection{Driven CTRW}

To determine $\mathrm{Var}[x(t)]_{\varepsilon}$ for the driven CTRW, we require the finite-time behavior of the first two moments of the jump number $N_t$. 
For a CTRW with a power-law waiting-time density $\psi(\tau)$ [Eq.~\eqref{eq:wt_dist_pow}], the process is a renewal process, and $N_t$ is the associated renewal counting process \cite{Cox}. 
Its statistics are therefore fully determined by the waiting-time distribution.

Let $T_r$ denote the sum of the first $r$ waiting times,
\begin{align}
  T_r=\sum_{j=1}^{r}\tau_j .
\end{align}
To characterize the statistics of the jump number $N_t$, we introduce the probability generating function
\begin{align}
  G(t,\zeta)=\Braket{\zeta^{N_t}}
  =\sum_{n=0}^{\infty}\zeta^n\,\Pr(N_t=n),
  \label{eq:def_G}
\end{align}
and its Laplace transform
\begin{align}
  \hat{G}(s,\zeta)=\int_{0}^{\infty}e^{-st}G(t,\zeta)\,dt .
  \label{eq:def_G_lap}
\end{align}
A standard renewal-theoretic expression gives \cite{Cox,Hou2018,Akimoto2018b}
\begin{align}
  \hat{G}(s,\zeta)
  =\frac{1}{s}+\frac{1}{s}\sum_{r=1}^{\infty}\zeta^{r-1}(\zeta-1)\hat{k}_r(s),
  \label{eq:hat{G}(s,zeta)}
\end{align}
where $\hat{k}_r(s)=\int_{0}^{\infty}k_r(t)e^{-st}\,dt$ is the Laplace transform of the PDF of $T_r$.
Since $T_r$ is a sum of IID waiting times, we have
$\hat k_r(s)=\hat\psi(s)^r$, with $\hat\psi(s)$ the Laplace transform of $\psi(\tau)$.
Substituting this into Eq.~\eqref{eq:hat{G}(s,zeta)} and summing the geometric series gives
\begin{equation}
\hat G(s,\zeta)=\frac{1-\hat\psi(s)}{s[1-\zeta\hat\psi(s)]}.
\end{equation}
This compact form highlights the renewal structure of the process: 
the finite-time statistics of $N_t$ are fully encoded in $\hat\psi(s)$. 
In particular, the first two moments of $N_t$ follow from derivatives of $\hat G(s,\zeta)$ at $\zeta=1$, which in turn determine the bias dependence of $\mathrm{Var}[x(t)]_{\varepsilon}$.

Differentiating $\hat{G}(s,\zeta)$ with respect to $\zeta$ and evaluating at $\zeta=1$ yields
\begin{align}
  \Braket{N_t}
  &=\mathcal{L}^{-1}\!\left\{\frac{1}{s}\frac{\hat{\psi}(s)}{1-\hat{\psi}(s)}\right\},
  \label{eq:N_t_psi}\\
  \Braket{N_t^2}
  &=\mathcal{L}^{-1}\!\left\{\frac{1}{s}\frac{\hat{\psi}(s)\,[1+\hat{\psi}(s)]}{[1-\hat{\psi}(s)]^2}\right\},
  \label{eq:N_t2_psi}
\end{align}
where $\mathcal{L}^{-1}$ denotes the inverse Laplace transform.
The Laplace-domain expressions \eqref{eq:N_t_psi} and \eqref{eq:N_t2_psi} are exact and encode the full time dependence of the jump-number statistics. 
In what follows, we extract their large-$t$ asymptotics using the small-$s$ expansion of $\hat\psi(s)$ together with Tauberian arguments \cite{Feller1971}. 
This allows us to determine the leading and subleading contributions to $\mathrm{Var}(N_t)$ and, via Eq.~\eqref{eq:VarXt_Nt}, to the variance $\mathrm{Var}[x(t)]_{\varepsilon}$.

\paragraph*{Case $\alpha>2$.}
When $\alpha>2$, the mean $\mu$ and variance $\sigma^2$ of $\tau$ are finite and
\begin{align}
  \hat{\psi}(s)\sim 1-\mu s+\frac{1}{2}(\mu^2+\sigma^2)s^2 .
  \label{eq:LTofPowalp>2}
\end{align}
This gives
\begin{align}
  \Braket{N_t}\sim \frac{t}{\mu},\qquad
  \mathrm{Var}(N_t)\sim \frac{\sigma^2}{\mu^3}t,
\end{align}
and hence
\begin{align}
\mathrm{Var}[x(t)]_{\varepsilon}
\sim
\left(
\frac{1}{\mu}
+
\frac{\sigma^2-\mu^2}{\mu^3}\,\varepsilon^2
\right)t.
  \label{eq:Var_CTRW_alpgt2}
\end{align}
Thus the response remains diffusive. Both terms scale linearly with time, so although the effective diffusion coefficient depends on $\varepsilon$, no competition between different asymptotic orders arises and hence no crossover occurs.

\paragraph*{Case $1<\alpha<2$.}
Here $\mu$ is finite but $\sigma^2$ diverges. The Laplace transform behaves as
\begin{align}
  \hat{\psi}(s)\sim 1-\mu s+b s^{\alpha},
  \label{eq:LTofPow1<alp<2}
\end{align}
where \(b\equiv |\Gamma(1-\alpha)|\tau_0^{\alpha}\).
This leads to
\begin{align}
    \braket{N_t} &\simeq \frac{t}{\mu} + \frac{1}{\mu^2} \cdot \frac{b}{\Gamma(3 - \alpha)} t^{2 - \alpha},
    \label{eq:fstmom_Nt_1<alp<2}
\end{align}
and
\begin{align}
  \mathrm{Var}(N_t)
  \sim \frac{2b}{\mu^3}\frac{\alpha-1}{\Gamma(4-\alpha)}t^{3-\alpha}.
  \label{eq:Var_Nt_1<alp<2}
\end{align}
Here we use ``$\sim$" to denote asymptotic equivalence at leading order,
and ``$\simeq$" to indicate asymptotic expansions retaining subleading terms.
Substituting into Eq.~\eqref{eq:VarXt_Nt} gives
\begin{align}
\mathrm{Var}[x(t)]_{\varepsilon}
\simeq
\frac{1-\varepsilon^2}{\mu}\, t
+
\frac{2b}{\mu^3}
\frac{\alpha-1}{\Gamma(4-\alpha)}
\varepsilon^2\, t^{3-\alpha}.
  \label{eq:Var_CTRW_alp12}
\end{align}
Since $3-\alpha>1$, the bias-induced contribution grows faster than the diffusive term.
Consequently, for $\varepsilon\neq0$ the long-time behavior becomes superdiffusive,
with $\mathrm{Var}[x(t)]_{\varepsilon}\sim t^{3-\alpha}$.
This implies a crossover from diffusive ($\propto t$) to superdiffusive ($\propto t^{3-\alpha}$) scaling.
The crossover time follows from equating the two contributions, indicating that weaker biases require longer observation times for the superdiffusive regime to become observable.

\paragraph*{Case $0<\alpha<1$.}
In this regime, $\mu$ diverges and
\begin{align}
  \hat{\psi}(s)\sim 1-a s^{\alpha},
  \label{eq:expansion_pow_dist}
\end{align}
where \(a\equiv \Gamma(1-\alpha)\tau_0^{\alpha}.\)
Then
\begin{align}
  \Braket{N_t}
  &\sim \frac{1}{a\Gamma(1+\alpha)}t^{\alpha},\label{eq:fstmom_Nt_0<alp<1}\\
  \mathrm{Var}(N_t)
  &\simeq \frac{1}{a^2}\left(\frac{2}{\Gamma(1+2\alpha)}-\frac{1}{\Gamma(1+\alpha)^2}\right)t^{2\alpha},
  \label{eq:Var_Nt_0<alp<1}
\end{align}
and Eq.~\eqref{eq:VarXt_Nt} yields
\begin{align}
  \mathrm{Var}\!\left[x(t)\right]_{\varepsilon}
  &\simeq \frac{1-\varepsilon^2}{a\Gamma(1+\alpha)}t^{\alpha}\nonumber\\
  &\quad+\frac{\varepsilon^2}{a^2}\left(\frac{2}{\Gamma(1+2\alpha)}-\frac{1}{\Gamma(1+\alpha)^2}\right)t^{2\alpha} .
  \label{eq:varofx(t)_CTRW}
\end{align}
The two contributions exhibit distinct scaling exponents, $t^{\alpha}$ and $t^{2\alpha}$.
For $\varepsilon\neq 0$, the bias-induced term grows faster and eventually dominates,
leading to a crossover from the unbiased subdiffusive scaling $\propto t^{\alpha}$ 
to a stronger subdiffusive behavior $\propto t^{2\alpha}$.
Equating the two contributions yields the crossover time
$t_c \sim \varepsilon^{-2/\alpha}$,
showing explicitly that weaker biases require longer observation times
for the bias-induced scaling to emerge.

\subsection{Driven QTM}

We now turn to the driven QTM. As in the CTRW case, the variance of the particle position can be expressed in terms of the mean and variance of the jump number $N_t$ via Eq.~\eqref{eq:VarXt_Nt}. The central task is therefore to determine the statistics of $N_t$ in the presence of quenched disorder.

In contrast to the CTRW, the waiting-time sequence in the QTM is not independent, since revisits to the same site reproduce the same trapping time. This induces correlations that complicate exact calculations. For $0<\alpha<1$, the variance of the driven QTM was previously derived by Burov \textit{et al.} using a different approach \cite{Burov2017}. Here we instead develop a jump-number-based framework that allows a unified treatment of driven QTM dynamics and extends naturally to the regime $\alpha>1$.

\subsubsection{Mean-field reduction of the waiting-time sum}

We first rewrite the sum of waiting times corresponding to $r$ jumps,
\begin{align}
  T_r=\sum_{j=1}^{r}\tau_j ,
\end{align}
in terms of site-based quantities. Here $\tau_j$ denotes the waiting time associated with the $j$-th jump. Let $N_{r,l}$ be the number of visits to site $l$ during $r$ steps, and let $\tilde{\tau}_l$ be the waiting time intrinsic to site $l$. Following Refs.~\cite{Miyaguchi2011,bouchaud90}, the waiting-time sum can be expressed as
\begin{align}
  T_r=\sum_{j=1}^{r}\tau_j=\sum_{l}N_{r,l}\tilde{\tau}_l .
\end{align}
This representation replaces the time-ordered waiting-time sequence $\{\tau_j\}$ with the spatially ordered set $\{\tilde{\tau}_l\}$. In the QTM, revisits to the same site reproduce the same trapping time, so the variables $\{\tau_j\}$ are generally correlated. By contrast, the site variables $\{\tilde{\tau}_l\}$ associated with different lattice sites are independent. As a result, the above representation expresses $T_r$ as a sum of IID random variables $\{\tilde{\tau}_l\}$ weighted by the visit numbers $N_{r,l}$.

Next we consider the effect of an external force. When the force biases the motion to the right, we assume that on sufficiently long time scales the visit number $N_{r,l}$ does not strongly depend on the site index $l$, and we approximate it in a mean-field manner as
\begin{align}
  N_{r,l}\sim\frac{r}{S_r},
\end{align}
where $S_r$ is the number of distinct sites visited within $r$ steps. Furthermore, following Ref.~\cite{Nagai1981}, we assume that in the regime of sufficiently strong bias, \(\Braket{S_r}\sim r\varepsilon\)
holds. Replacing $S_r$ by its mean under this approximation yields
\begin{align}
  N_{r,l}\sim\frac{1}{\varepsilon}.
\end{align}
Here, we employ a mean-field-type approximation, assuming that the number of visits is approximately uniformly distributed across the visited sites. 
This type of approximation is commonly used in the analysis of the QTM even in the absence of bias, where it effectively captures the scaling behavior of observables. 
In the present case, the presence of a bias further promotes this uniformity by suppressing backward motion, thereby providing additional justification for the approximation.

With these assumptions, the waiting-time sum is approximated as
\begin{align}
  T_r\sim \frac{1}{\varepsilon}\sum_{l=1}^{r\varepsilon}\tilde{\tau}_l .
  \label{eq:T_r}
\end{align}
This mean-field approximation is intended to capture the leading $\varepsilon$ dependence of the renewal statistics in the driven regime where the particle visits many distinct sites, $r\varepsilon\gg 1$.
In what follows, we use it to obtain the asymptotic scaling of the bias-induced contribution to $\mathrm{Var}[x(t)]_\varepsilon$.

\subsubsection{Moments of the jump number}

The procedure for evaluating the moments of $N_t$ is the same as for the CTRW. We therefore begin by determining the Laplace transform $\hat{k}_r(s)$ of the PDF $k_r(t)$ of $T_r$.  From Eq.~\eqref{eq:T_r}, $\hat{k}_r(s)$ is approximated as
\begin{align}
  \hat{k}_r(s)\approx \left\{\hat{\phi}_{\varepsilon}(s)\right\}^{r\varepsilon},
\end{align}
where we defined
\begin{align}
  \hat{\phi}_{\varepsilon}(s)
  \equiv \Braket{e^{-s\tilde{\tau}/\varepsilon}}
  =\int_{0}^{\infty}e^{-s\tilde\tau/\varepsilon}\psi(\tilde\tau)\,d\tilde\tau ,
  \label{eq:phi(s)}
\end{align}
and $\psi(\tilde{\tau})$ is the PDF of the independent site waiting time $\tilde{\tau}$.

Using this approximation for $\hat{k}_r(s)$ into Eq.~\eqref{eq:hat{G}(s,zeta)} and differentiating with respect to $\zeta$ at $\zeta=1$, we obtain the first and second moments of the renewal number in the driven QTM as
\begin{align}
  \Braket{N_t}_{\varepsilon>0}
  &=\mathcal{L}^{-1}\left\{\frac{1}{s}\frac{\left\{\hat{\phi}_{\varepsilon}(s)\right\}^{\varepsilon}}{1-\left\{\hat{\phi}_{\varepsilon}(s)\right\}^{\varepsilon}}\right\},
  \label{eq:fstmom_Nt_psi_eps}\\
  \Braket{N_t^2}_{\varepsilon>0}
  &=\mathcal{L}^{-1}\left\{\frac{1}{s}\frac{\left\{\hat{\phi}_{\varepsilon}(s)\right\}^{\varepsilon}\left[1+\left\{\hat{\phi}_{\varepsilon}(s)\right\}^{\varepsilon}\right]}{\left[1-\left\{\hat{\phi}_{\varepsilon}(s)\right\}^{\varepsilon}\right]^2}\right\}.
  \label{eq:sndmom_Nt_psi_eps}
\end{align}
Since the force dependence of the renewal statistics is essential in the QTM, we explicitly indicate the $\varepsilon$ dependence of these moments. In the following, we specify the asymptotic form of $\hat{\phi}_{\varepsilon}(s)$ as $s\to 0$ from the concrete form of $\psi(\tilde{\tau})$, and evaluate the renewal-number moments for the driven QTM by using Eqs.~\eqref{eq:fstmom_Nt_psi_eps} and \eqref{eq:sndmom_Nt_psi_eps}.

\subsubsection{Variance of the displacement}

\paragraph*{Case $\alpha>2$.}
We first consider the case $\alpha>2$. In this regime, the first and second moments of the site waiting time $\tilde{\tau}$ are finite, so that $\hat{\phi}_{\varepsilon}(s)$ can be expanded as $s\to 0$:
\begin{align}
  \hat{\phi}_{\varepsilon}(s)
  \sim 1-\mu\frac{s}{\varepsilon}
  +\frac{1}{2}(\sigma^2+\mu^2)\frac{s^2}{\varepsilon^2},
  \label{eq:phis_alp>2}
\end{align}
where $\mu\equiv\Braket{\tilde{\tau}}$ and $\sigma^2\equiv\mathrm{Var}(\tilde{\tau})$.
For $\alpha>2$, the second moment of the waiting time is finite and the coefficients of the variance depend explicitly on these moments. Therefore, the moments $\mu$ and $\sigma^2$ must be evaluated from the underlying energy landscape of the QTM rather than approximating the waiting-time distribution by a simple power-law form.

The trap depth $E$ is exponentially distributed with mean $k_{\rm B}T_{\rm g}$,
\begin{align}
  \rho(E)=\frac{1}{k_{\rm B}T_{\rm g}}\exp\!\left(-\frac{E}{k_{\rm B}T_{\rm g}}\right).
\end{align}
The waiting time at a site follows the Arrhenius law
$\tilde{\tau}=\tau_0 e^{E/(k_{\rm B}T)}$.
Thus the conditional distribution of $\tilde{\tau}$ given $E$ is exponential with mean $\tau_0 e^{E/(k_{\rm B}T)}$,
\begin{align}
  \rho(\tilde{\tau}\mid E)
  =\frac{e^{-E/(k_{\rm B}T)}}{\tau_0}
  \exp\!\left[-\frac{\tilde{\tau}}{\tau_0}e^{-E/(k_{\rm B}T)}\right].
\end{align}
Marginalizing over $E$ gives
\begin{align}
  \rho(\tilde{\tau})
  &=\frac{1}{k_{\rm B}T_{\rm g}\tau_0}
  \int_{0}^{\infty}
  e^{-\frac{E}{k_{\rm B}T_{\rm g}}-\frac{E}{k_{\rm B}T}}
  \exp\!\left[-\frac{\tilde{\tau}}{\tau_0}e^{-E/(k_{\rm B}T)}\right]\,dE .
\end{align}
Introducing $\alpha\equiv T/T_{\rm g}$ and changing variables as $u=e^{-E/(k_{\rm B}T)}$, we obtain
\begin{align}
  \rho(\tilde{\tau})
  =\frac{\alpha}{\tau_0}\int_{0}^{1}u^{\alpha} \exp\!\left(-\frac{\tilde{\tau}}{\tau_0}u\right)\,du .
\end{align}
From this expression, the mean and variance of \(\tilde \tau\) are given by
\begin{align}
  \mu &= \tau_0\frac{\alpha}{\alpha-1},\\
  \sigma^{2} &= \tau_0^{2}\left[\frac{2\alpha}{\alpha-2}-\left(\frac{\alpha}{\alpha-1}\right)^{2}\right].
\end{align}
A noteworthy point is that $\mu$ coincides with the first moment of the power-law approximation,
$\tau_0\,\alpha/(\alpha-1)$, whereas $\sigma^2$ does not in general. Hence, for $\alpha>2$ one cannot
simply treat the independent site waiting time $\tilde{\tau}$ as a power-law random variable when discussing the
coefficients of the variance.

Substituting Eq.~\eqref{eq:phis_alp>2} into Eqs.~\eqref{eq:fstmom_Nt_psi_eps} and \eqref{eq:sndmom_Nt_psi_eps}, we obtain the first and second moments of the renewal number as
\begin{align}
  \Braket{N_t}_{\varepsilon>0}
  &\sim \frac{t}{\mu}
  +\frac{\sigma^2-\mu^2\varepsilon}{2\mu^2\varepsilon},
  \label{eq:fstmom_Nt_Q_alp>2}\\
  \Braket{N_t^2}_{\varepsilon>0}
  &\sim \frac{t^2}{\mu^2}
  +\frac{2\sigma^2-\mu^2\varepsilon}{\mu^3\varepsilon}t .
  \label{eq:sndmom_Nt_Q_alp>2}
\end{align}
Hence, the variance of the renewal number behaves as
\begin{align}
  \mathrm{Var}(N_t)_{\varepsilon>0}
  \sim \frac{\sigma^2}{\mu^3}\frac{t}{\varepsilon}.
  \label{eq:VarofNt_Q}
\end{align}
In contrast to the CTRW, where the renewal statistics is independent of the force, the QTM exhibits an explicit force dependence, $\mathrm{Var}(N_t)\propto \varepsilon^{-1}$. This reflects that a stronger bias increases the rate of reaching new sites, thereby weakening the effective correlations induced by revisits. Indeed, taking $\varepsilon\to 1$ in Eqs.~\eqref{eq:fstmom_Nt_Q_alp>2}--\eqref{eq:VarofNt_Q} recovers expressions of the same form as in the CTRW, where waiting-time correlations are absent.

Finally, combining Eq.~\eqref{eq:VarXt_Nt} with the above results, we obtain the asymptotic variance of the particle position,
\begin{align}
  \mathrm{Var}[x(t)]_{\varepsilon>0}
  \sim \left(\frac{1}{\mu}
  +\frac{\sigma^2 \varepsilon}{\mu^3} 
  -\frac{\varepsilon^{2}}{\mu} \right)t.
  \label{eq:Var_QTM_gt2}
\end{align}
All terms scale linearly with time, indicating that the long-time transport remains diffusive. 
In contrast to the CTRW case, the variance contains a contribution proportional to $\varepsilon t$. 
This term originates from the quenched spatial disorder of the trap landscape and reflects the coupling between the bias and the statistics of trapping times.

\paragraph*{Case $1<\alpha<2$.} Next we consider the case $1<\alpha<2$. In this regime, the mean waiting time is finite while the second moment diverges, and thus $\hat{\phi}_{\varepsilon}(s)$ admits the small-$s$ expansion
\begin{align}
  \hat{\phi}_{\varepsilon}(s)
  \sim 1-\mu\frac{s}{\varepsilon}+b\left(\frac{s}{\varepsilon}\right)^{\alpha},
  \label{eq:LTofPow1<alp<2Q}
\end{align}
where $\mu\equiv\Braket{\tilde{\tau}}$ and $b\equiv \tau_0^{\alpha}|\Gamma(1-\alpha)|$ is a constant. Substituting Eq.~\eqref{eq:LTofPow1<alp<2Q} into Eqs.~\eqref{eq:fstmom_Nt_psi_eps} and \eqref{eq:sndmom_Nt_psi_eps}, we obtain the first and second moments of the renewal number as
\begin{align}
  \Braket{N_t}_{\varepsilon>0}
  &\sim \frac{t}{\mu}
  +\frac{1}{\mu^2}\frac{b}{\Gamma(3-\alpha)}\varepsilon^{1-\alpha}t^{2-\alpha},\label{eq:fstmom_Nt_Q_1<alp<2}\\
  \Braket{N_t^2}_{\varepsilon>0}
  &\sim \frac{t^2}{\mu^2}
  +\frac{4}{\mu^3}\frac{b}{\Gamma(4-\alpha)}\varepsilon^{1-\alpha}t^{3-\alpha}.
\end{align}
Therefore, the asymptotic variance of the renewal number is given by
\begin{align}
  \mathrm{Var}(N_t)_{\varepsilon>0}
  \sim \frac{2b}{\mu^3}\frac{\alpha-1}{\Gamma(4-\alpha)}\varepsilon^{1-\alpha}t^{3-\alpha}.
  \label{eq:Var_Nt_Q_1<alp<2}
\end{align}
Substituting these results into Eq.~\eqref{eq:VarXt_Nt}, we obtain the asymptotic variance of the particle position,
\begin{align}
  \mathrm{Var}[x(t)]_\varepsilon \sim \frac{t}{\mu}(1-\varepsilon^2) + \frac{2b}{\mu^3} \frac{\alpha-1}{\Gamma(4 - \alpha)} \varepsilon^{3 - \alpha} t^{3 - \alpha}.
  \label{eq:QTM_Varxt_alp12}
\end{align}
This expression provides the scaling of $\mathrm{Var}[x(t)]_{\varepsilon>0}$ with respect to $t$ and $\varepsilon$ in the long-time limit $t\to\infty$ for $1<\alpha<2$. As in the CTRW result [Eq.~\eqref{eq:Var_CTRW_alp12}], the variance in the QTM also grows as $\propto t^{3-\alpha}$, indicating enhanced diffusion for $1<\alpha<2$. However, the force dependence differs: while the CTRW response scales as $\propto \varepsilon^{2}$, the driven QTM exhibits $\propto \varepsilon^{3-\alpha}$. This difference originates from the quenched (spatially frozen) traps in the QTM, which render the renewal-number variance force dependent.

\paragraph*{Case $0<\alpha<1$.}Finally, we consider the case $0<\alpha<1$. Although the result in this regime has been derived by Burov \textit{et al.}\ using a different method \cite{Burov2017}, we here apply the same procedure as for $\alpha>1$ and confirm that we recover the same asymptotic behavior.

Since the mean waiting time diverges in this regime, $\hat{\phi}_{\varepsilon}(s)$ behaves for $s\to 0$ as
\begin{align}
  \hat{\phi}_{\varepsilon}(s)
  = 1-a\left(\frac{s}{\varepsilon}\right)^{\alpha}+o(s^{\alpha}),
  \label{eq:LTofPow0<alp<1Q}
\end{align}
where $a\equiv \Gamma(1-\alpha)\tau_0^{\alpha}$ 
is a constant, different from $b$ appearing in the case $1<\alpha<2$. Substituting Eq.~\eqref{eq:LTofPow0<alp<1Q} into Eqs.~\eqref{eq:fstmom_Nt_psi_eps} and \eqref{eq:sndmom_Nt_psi_eps}, we obtain
\begin{align}
  \Braket{N_t}_{\varepsilon>0}
  &\sim \frac{1}{a\Gamma(1+\alpha)}t^{\alpha}\varepsilon^{-(1-\alpha)},
  \label{eq:fstmom_Nt_0<alp<1Q}\\
  \Braket{N_t^2}_{\varepsilon>0}
  &\sim \frac{2}{a^{2}\Gamma(1+2\alpha)}t^{2\alpha}\varepsilon^{-2(1-\alpha)}.
\end{align}
Hence,
\begin{align}
  \mathrm{Var}(N_t)_{\varepsilon>0}
  \sim \frac{1}{a^{2}}
  \left(
    \frac{2}{\Gamma(1+2\alpha)}
    -\frac{1}{\Gamma(1+\alpha)^{2}}
  \right)
  t^{2\alpha}\varepsilon^{2\alpha-2}.\label{eq:Var_Nt_0<alp<1Q}
\end{align}
Substituting this into Eq.~\eqref{eq:VarXt_Nt}, we obtain the asymptotic variance of the displacement,
\begin{align}
  \mathrm{Var}[x(t)]_{\varepsilon>0}
  \sim \frac{1}{a^{2}}
  \left(
    \frac{2}{\Gamma(1+2\alpha)}
    -\frac{1}{\Gamma(1+\alpha)^{2}}
  \right)
  t^{2\alpha}\varepsilon^{2\alpha}.
  \label{eq:varofx(t)_QTM}
\end{align}
Although our derivation assumes $r\varepsilon\gg 1$ when approximating the jump number $r$ and the force parameter $\varepsilon$, Eq.~\eqref{eq:varofx(t)_QTM} agrees (including the prefactor) with the $\varepsilon\to 0$ result by Burov \textit{et al.} \cite{Burov2017}. A similar asymptotic behavior in the strongly heterogeneous limit $\alpha\to 0$ is also known to be obtained by a renormalization-group method \cite{Monthus2004}.

\section{Observation-time-induced crossover}

To quantify the effect of the bias on transport fluctuations, we introduce the normalized transport response
\begin{align}
  R(t,\varepsilon)=\frac{\mathrm{Var}[x(t)]_{\varepsilon>0}}{\mathrm{Var}[x(t)]_{\varepsilon=0}} .
\end{align}
This quantity measures how the external bias modifies the magnitude of transport fluctuations relative to the unbiased dynamics.

\begin{figure*}[t]
    \centering
    \includegraphics[width=\linewidth]{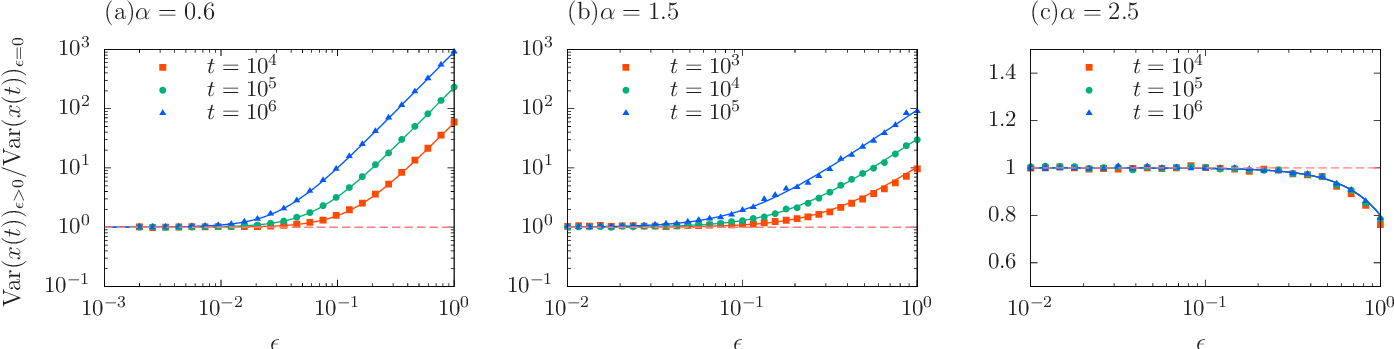}
    \caption{Normalized transport response of the CTRW for different observation times. 
(a) $\alpha=0.6$, (b) $\alpha=1.5$, and (c) $\alpha=2.5$. 
The solid lines show the asymptotic behaviors of the variance under an external force, obtained Eqs.~\eqref{eq:varofx(t)_CTRW}, \eqref{eq:Var_CTRW_alp12}, and \eqref{eq:Var_CTRW_alpgt2}. The dashed horizontal line indicates $R=1$ as a reference.
}
    \label{fig:OTIC_CTRW}
\end{figure*}

\begin{figure*}[t]
    \centering
    \includegraphics[width=\linewidth]{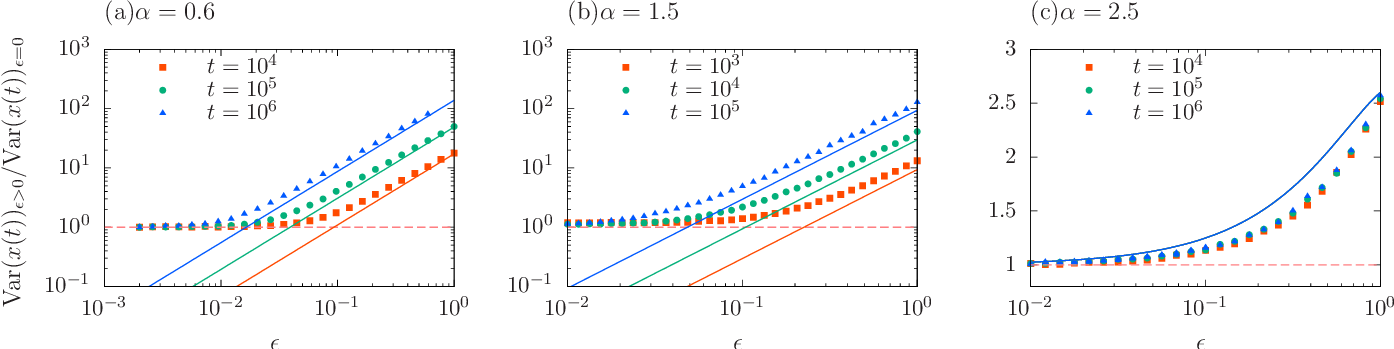}
    \caption{Normalized transport response of the QTM for different observation times. 
(a) $\alpha=0.6$, (b) $\alpha=1.5$, and (c) $\alpha=2.5$. 
For $\alpha<2$, the solid lines represent the leading-order contribution 
to the displacement variance in the large-bias limit, corresponding to 
the second term of Eq.~\eqref{eq:VarXt_Nt}, $\left(\mathrm{Var}(N_t)_{\varepsilon>0}-\Braket{N_t}_{\varepsilon>0}\right)\varepsilon^{2}$.
For $\alpha>2$, where an observation-time induced crossover does not occur, the solid lines depict the asymptotic expression valid over the entire $\varepsilon$ range, given by Eq.~(\ref{eq:Var_QTM_gt2}). The dashed horizontal line indicates $R=1$ as a reference.
}
    \label{fig:OTIC_QTM}
\end{figure*}

Figures~\ref{fig:OTIC_CTRW} and \ref{fig:OTIC_QTM} show the simulation results for the normalized response $R(t,\varepsilon)$ in the CTRW and the QTM, respectively, for several values of the exponent $\alpha$.
For the CTRW (Fig.~\ref{fig:OTIC_CTRW}), the theoretical predictions agree well with the simulation results.
For the QTM (Fig.~\ref{fig:OTIC_QTM}), the theoretical curves asymptotically agree with the simulations in the regime $0<\alpha<1$ for sufficiently large $\varepsilon$.
For $\alpha>2$, small quantitative deviations between theory and simulations are observed. 
These deviations originate from the mean-field approximation used in the theoretical analysis, 
where the visit number was approximated as $N_{r,l}\sim 1/\varepsilon$. 
This approximation captures the leading dependence on the bias but neglects detailed fluctuations in the number of revisits. 
Notably, the approximation becomes exact in the limit $\varepsilon=1$, where the particle moves strictly in one direction and revisits do not occur.

Focusing on the $\varepsilon$ dependence of the response, we first note a common trend in both models.
For sufficiently weak forcing, $R(t,\varepsilon)\simeq 1$, indicating that the fluctuations appear insensitive to the external force.
As the force increases, $R(t,\varepsilon)$ deviates from unity, showing that the fluctuations respond to the driving.
For $\alpha<2$, the force scale at which the response sets in depends on the observation time: the longer the observation time $t$, the smaller the force at which $R(t,\varepsilon)$ departs from unity.
Consequently, the crossover point between the apparently unresponsive and responsive regimes shifts with the observation time, demonstrating an observation-time-induced crossover.

This crossover originates from the fact that, once an external force is applied, the system is driven out of equilibrium and a bias-induced scaling of the fluctuations emerges. 
More precisely, the variance of the displacement in Eq.~\eqref{eq:VarXt_Nt} crosses over from the scaling of the unbiased motion encoded in the first term, $\Braket{N_t}$, to the bias-induced scaling governed by the second term. 
In addition to the presence or absence of the force dependence, the two contributions generally exhibit different scalings with the observation time $t$. 
This difference in the $t$-scalings leads to an observation-time-induced crossover in the force response.
For both the driven CTRW and the driven QTM, no observation-time-induced crossover occurs for $\alpha>2$.
In this regime, the second moment of the waiting time is finite, and both $\Braket{N_t}$ and $\mathrm{Var}(N_t)$ scale linearly with the observation time.
Consequently, the two terms in Eq.~\eqref{eq:VarXt_Nt} share the same $t$ scaling for any bias strength $\varepsilon$, so the crossover threshold does not depend on $t$.

\section{Observation-time scaling of the response bias}

To estimate the crossover boundary more explicitly, we compare the two contributions to the variance appearing in Eq.~\eqref{eq:VarXt_Nt}. 
The first term becomes relevant only when the bias-induced contribution is sufficiently small, in which case it can be approximated by the renewal function of the unbiased process, $\Braket{N_t}_{\varepsilon=0}$. 
Thus,
\begin{align}
  \mathrm{Var}\!\left[x(t)\right]_{\varepsilon}
  \simeq \Braket{N_t}_{\varepsilon=0}
  +\left(\mathrm{Var}(N_t)_{\varepsilon>0}-\Braket{N_t}_{\varepsilon>0}\right)\varepsilon^{2}.
\end{align}
The crossover boundary can therefore be estimated by balancing the leading-order asymptotic behaviors of $\Braket{N_t}_{\varepsilon=0}$ and $\left(\mathrm{Var}(N_t)_{\varepsilon}-\Braket{N_t}_{\varepsilon}\right)\varepsilon^{2}$.

\subsection{Driven CTRW}

For the CTRW, the renewal statistics is independent of the bias $\varepsilon$. Hence, in the regime $\alpha<2$ where the variance exhibits the observation-time-induced crossover, the observation-time dependence of the crossover (threshold) bias can be obtained by equating the leading-order contributions of $\Braket{N_t}$ and $\left(\mathrm{Var}(N_t)-\Braket{N_t}\right)\varepsilon^{2}$.
Using Eqs.~\eqref{eq:fstmom_Nt_1<alp<2} and \eqref{eq:Var_Nt_1<alp<2}, as well as Eqs.~\eqref{eq:fstmom_Nt_0<alp<1} and \eqref{eq:Var_Nt_0<alp<1}, the observation-time dependence of the crossover (threshold) bias $\varepsilon_c$ is obtained as
\begin{align}
  \varepsilon_c(t)=
  \begin{cases}
    \displaystyle
    \sqrt{
      \frac{a \Gamma(1+\alpha)\Gamma(2\alpha +1)}
      {2\Gamma(1+\alpha)^2-\Gamma(2\alpha+1)}
    }\;t^{-\frac{\alpha}{2}}
    & (0<\alpha<1),\\[8pt]
    \\
    \displaystyle
    \sqrt{\frac{\mu^2}{2b}\frac{\Gamma(4-\alpha)}{\alpha-1}}\;t^{-(2-\alpha)/2}
    & (1<\alpha<2).
  \end{cases}
  \label{eq:DCepsCTRW}
\end{align}
In both $\alpha$ regimes, $\varepsilon_c(t)$ is a decreasing function of $t$, meaning that a longer observation makes the bias-induced contribution visible even for weaker bias. 
This behavior reflects the competition between the unbiased fluctuation term $\Braket{N_t}$ and the bias-induced contribution $\left(\mathrm{Var}(N_t)-\Braket{N_t}\right)\varepsilon^{2}$. 
Because the latter grows faster with time than the former when $\alpha<2$, the bias-induced contribution eventually dominates even for small $\varepsilon$. 
Consequently, the force scale required to observe the response decreases algebraically with the observation time.

\subsection{Driven QTM}

For the QTM, we use the known results for the mean renewal number in the unbiased case \cite{bouchaud90, PhysRevE.86.041137, Miyaguchi2011, PhysRevE.67.026128}:
\begin{align}
  \Braket{N_t}_{\varepsilon=0}\sim
  \begin{cases}
    C_\alpha\, t^{\frac{2\alpha}{1+\alpha}} & (0<\alpha<1),\\
    t/\mu & (1<\alpha<2),
  \end{cases}
\end{align}
where $C_\alpha$ is an $\alpha$-dependent constant. To date, its exact closed-form expression is not known except for special cases such as $\alpha\to 0$ and $\alpha\to 1$\cite{PhysRevLett.106.140602}.

Using the moments of the renewal number under bias, Eqs.~\eqref{eq:fstmom_Nt_Q_1<alp<2} and \eqref{eq:Var_Nt_Q_1<alp<2}, as well as Eqs.~\eqref{eq:fstmom_Nt_0<alp<1Q} and \eqref{eq:Var_Nt_0<alp<1Q}, we evaluate the asymptotic behavior of the bias-induced contribution,
$\left(\mathrm{Var}(N_t)_{\varepsilon>0}-\Braket{N_t}_{\varepsilon>0}\right)\varepsilon^{2}$,
and compare it with the unbiased scaling. This yields the crossover (threshold) bias in the QTM as 
\begin{align}
\varepsilon_c(t)\sim 
\begin{cases}
A_\alpha\, t^{-\frac{\alpha}{1+\alpha}}
& (0<\alpha<1),\\[6pt]
B_\alpha\, t^{-\frac{2-\alpha}{3-\alpha}}
& (1<\alpha<2),
\end{cases}
\label{eq:e_t_QTM}
\end{align}
where
\begin{align}
&A_\alpha=
\left[
\frac{C_\alpha a^2 \Gamma(1+2\alpha)\Gamma(1+\alpha)^2}
{2\Gamma(1+\alpha)^2-\Gamma(1+2\alpha)}
\right]^{\frac{1}{2\alpha}},\\
&B_\alpha=    \displaystyle
    \left(
      \frac{\mu^2}{2b}\frac{\Gamma(4-\alpha)}{\alpha-1}
    \right)^{\frac{1}{3-\alpha}}.
\end{align}
Equation~\eqref{eq:e_t_QTM} is obtained by matching the unbiased scaling of $\Braket{N_t}_{\varepsilon=0}$ with the asymptotic driven contribution extracted from the mean-field renewal statistics.
Accordingly, $\varepsilon_c(t)$ should be viewed as a crossover scale associated with a bias-controlled relaxation time toward the driven nonequilibrium scaling.
Interestingly, in the strongly heterogeneous regime $0<\alpha<1$, the crossover scale in the variance obeys the same time dependence as the well-known crossover in the mean displacement from linear to nonlinear response, $\varepsilon_c(t)\propto t^{-\alpha/(1+\alpha)}$\cite{bouchaud90,PhysRevE.67.065105}.

As in the CTRW case, the threshold bias $\varepsilon_c(t)$ in the QTM is a decreasing function of the observation time $t$. 
However, even within the same $\alpha$ range, the corresponding power-law exponents differ between the two models. 
This difference highlights the impact of the quenched environment on the response of the variance under bias, i.e., on the nonequilibrium fluctuations induced by driving.

\subsection{Role of Quenched Heterogeneity}

We now examine how quenched disorder affects the observation-time scaling of the response threshold.
In particular, we compare the scaling of the threshold bias between the CTRW and the QTM.
To this end, we introduce the scaling exponent $\nu$ defined by
$\varepsilon_c\propto t^{-\nu}$.

\begin{figure}[tb]
    \centering
    \includegraphics[width=\linewidth]{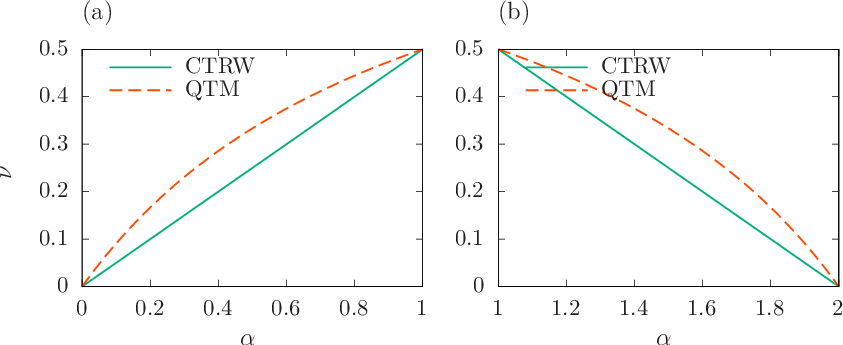}
    \caption{Scaling exponent $\nu$ of the threshold bias $\varepsilon_c$ for the driven CTRW and the driven QTM.
    (a) $0<\alpha<1$, (b) $1<\alpha<2$. The solid and dashed lines correspond to the scaling exponents for the driven CTRW and the driven QTM, respectively.}
    \label{fig:exponent}
\end{figure}

Figure~\ref{fig:exponent} summarizes the scaling exponent $\nu$ for the driven CTRW and the driven QTM.
As is evident from the figure, in both regimes $0<\alpha<1$ and $1<\alpha<2$, the exponent $\nu$ is always larger for the QTM than for the CTRW.
This implies that, for the same observation time and the same value of $\alpha$, the QTM exhibits a detectable response at a smaller bias than the CTRW.
This can be understood qualitatively as follows. In the QTM, each site is assigned a quenched mean waiting time, and upon revisiting the same site the waiting time is drawn again from the same distribution, which induces correlations in the sequence of waiting times along a trajectory. In contrast to the CTRW, where waiting times are independently renewed at every step, the effective short-time variability is reduced and encounters with extremely long waiting times are relatively suppressed. As a result, the particle is less affected by rare long trapping events and more readily crosses over to the biased asymptotic regime at shorter observation times, i.e., at smaller bias. This provides a qualitative origin of why the QTM shows a detectable response at smaller bias (larger $\nu$) than the CTRW for the same $\alpha$.

\begin{figure}[tb]
  \centering
  \includegraphics[width=0.49\columnwidth]{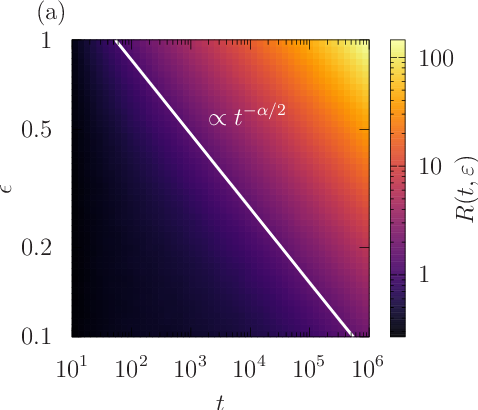}\hfill
  \includegraphics[width=0.49\columnwidth]{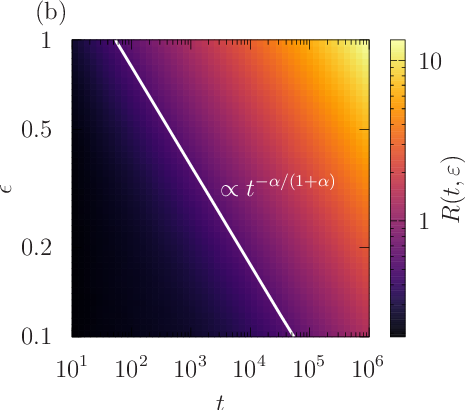}
  \caption{Heatmap of the normalized transport response for (a) the driven CTRW and (b) the driven QTM.
The solid lines indicate the theoretical crossover threshold
given by Eq.~\eqref{eq:DCepsCTRW} for the CTRW and Eq.~\eqref{eq:e_t_QTM} for the QTM.
The coefficient $C_\alpha$ is obtained numerically.
  }
  \label{fig:var_heatmap}
\end{figure}

Figure~\ref{fig:var_heatmap} shows a heat map of the normalized transport response obtained from simulations in the $(\varepsilon,t)$ plane. 
For both models, the variance can be clearly separated into two regions: one where it is essentially insensitive to the bias and another where it changes with the bias. 
The boundary between these regions is described by Eqs.~\eqref{eq:DCepsCTRW} and \eqref{eq:e_t_QTM} for the CTRW and the QTM, respectively, and is plotted as a white solid line in the figure. 
(For Eq.~\eqref{eq:e_t_QTM}, the coefficient $C_\alpha$ is taken from the value estimated numerically from the simulations.) These results indicate that the bias strength required to induce a transport response depends continuously on the observation time $t$, and that the associated boundary arises from a change in the dominant scaling of the variance: from the unbiased behavior, $\mathrm{Var}[x(t)]_{\varepsilon=0}\sim \Braket{N_t}_{\varepsilon=0}$, to the bias-induced behavior,
$\mathrm{Var}[x(t)]_{\varepsilon>0}\sim \left(\mathrm{Var}(N_t)_{\varepsilon>0}-\Braket{N_t}_{\varepsilon>0}\right)\varepsilon^{2}$.

\section{Mechanism: Bias-Dependent Relaxation of the Transport Response}

\begin{figure*}[tb]
    \centering
    \includegraphics[width=\linewidth]{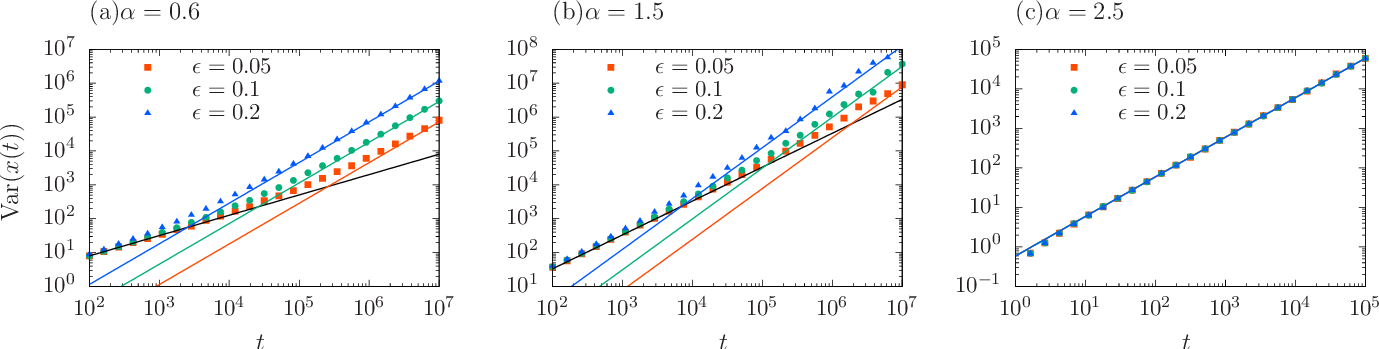}
    \caption{Convergence of the variance under bias, $\mathrm{Var}[x(t)]_{\varepsilon>0}$, to the nonequilibrium scaling in the CTRW. The black solid line represents the asymptotic behavior of the unbiased variance, $\mathrm{Var}[x(t)]_{\varepsilon=0}$, whereas the other solid lines represent the asymptotic behaviors of the biased variance, $\mathrm{Var}[x(t)]_{\varepsilon>0}$, for each value of $\varepsilon$.
    }
    \label{fig:CTRW_t_resp}
\end{figure*}

\begin{figure*}[tb]
    \centering
    \includegraphics[width=\linewidth]{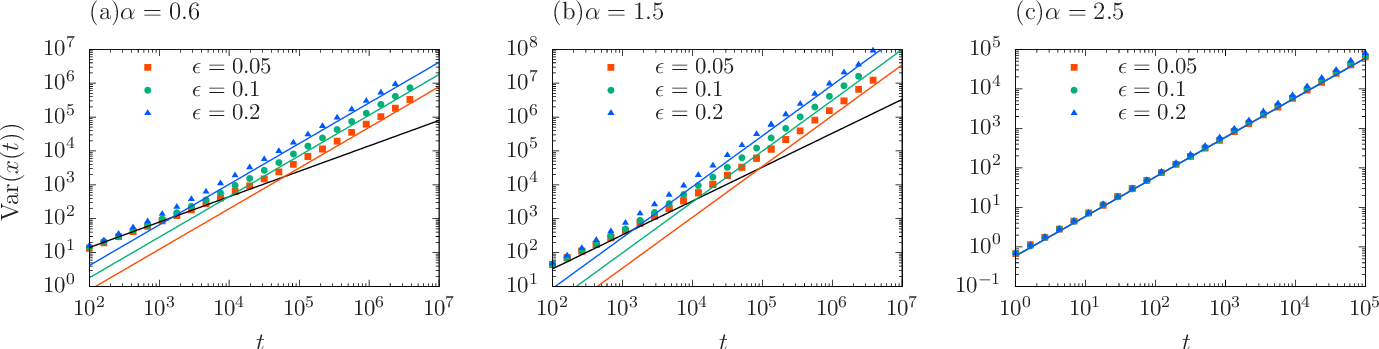}
    \caption{Convergence of the variance under bias, $\mathrm{Var}[x(t)]_{\varepsilon>0}$, to the nonequilibrium scaling in the QTM. The black solid line represents the asymptotic behavior of the unbiased variance, $\mathrm{Var}[x(t)]_{\varepsilon=0}$, whereas the other solid lines represent the asymptotic behaviors of the biased variance, $\mathrm{Var}[x(t)]_{\varepsilon>0}$, for each value of $\varepsilon$.}
    \label{fig:QTM_t_resp}
\end{figure*}

To understand the origin of the observation-time-induced crossover
from the viewpoint of the relaxation timescale of the transport response,
we examine the time evolution of the variance of the displacement.
Figures~\ref{fig:CTRW_t_resp} and \ref{fig:QTM_t_resp} show simulation results for the time evolution of the displacement variance
for the driven CTRW and the driven QTM, respectively.
The black solid line represents the asymptotic behavior of the unbiased variance,
$\mathrm{Var}[x(t)]_{\varepsilon=0}$, whereas the other solid lines represent
the asymptotic behaviors of the biased variance,
$\mathrm{Var}[x(t)]_{\varepsilon>0}$, for different values of $\varepsilon$.

For $\alpha<2$, where the observation-time-induced crossover occurs,
the variance does not immediately follow the bias-induced scaling even when
a bias is applied.
Instead, it approaches the asymptotic nonequilibrium scaling only after a
sufficiently long time.
Moreover, the time required to reach this bias-induced scaling depends
strongly on the bias strength $\varepsilon$.
This behavior suggests the presence of a characteristic timescale.
We define the transport response time \(t_{\rm resp}\) as the
$\varepsilon$-dependent timescale required for the variance to converge
to the bias-induced scaling.

\begin{figure}[tb]
    \centering
    \includegraphics[width=\linewidth]{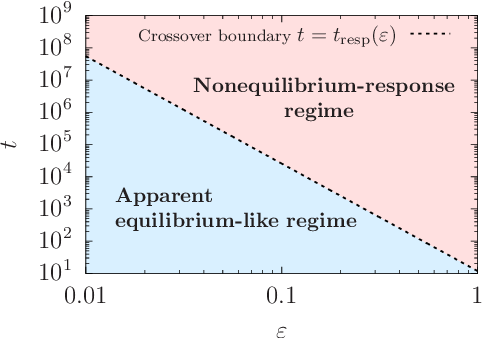}
    \caption{Transport response time $t_{\mathrm{resp}}(\varepsilon)=\varepsilon_c^{-1}(\varepsilon)$ as a function of the bias strength $\varepsilon$. The dashed line represents the transport response time for the driven CTRW ($\alpha=0.6$). The blue and red shaded regions indicate the \emph{apparent equilibrium-like regime}
($t<t_{\mathrm{resp}}$) and the \emph{detectable nonequilibrium-response regime}
($t>t_{\mathrm{resp}}$), respectively.
}
    \label{fig:t_eps_plane}
\end{figure}

The transport response time as a function of the bias strength $\varepsilon$ can then be introduced as the inverse  function of $\varepsilon_c(t)$:
\begin{align}
  t_{\rm resp}(\varepsilon)\equiv \varepsilon_c^{-1}(\varepsilon).
\end{align}
In the $t$--$\varepsilon$ plane, the transport response time $t_{\mathrm{resp}}(\varepsilon)$ is shown in Fig.~\ref{fig:t_eps_plane}.
The boundary $t=t_{\mathrm{resp}}(\varepsilon)$ (dashed line) separates the parameter region where the bias-induced response is not yet observable from the region where the nonequilibrium response becomes detectable.
When the observation time $t$ exceeds this boundary, the dynamics crosses over from an \emph{apparent equilibrium-like regime} to a \emph{detectable nonequilibrium response}.
Accordingly, the threshold bias $\varepsilon_c(t)$, defined by the intersection with this boundary, shifts as the observation time is varied.
This observation-time-induced crossover therefore requires that the transport response time depends on an external control parameter, here the driving bias $\varepsilon$.

\section{Discussion}

The present results suggest that observation-time-induced crossover is not specific to CTRW and QTM, but may be a more general consequence of slow relaxation. From this perspective, the relevant issue is not only the magnitude of the external control parameter, but also whether the observation window is long enough for the system to approach its asymptotic response. A useful interpretation is therefore to view the crossover as a competition between the observation time and a control-parameter-dependent relaxation timescale. Although the microscopic origins of such slow relaxation differ across systems, a common feature is that finite-time measurements can mask the asymptotic response and thereby shift the apparent crossover point.

Finite-time effects of the mean displacement in the driven QTM have also been reported \cite{bouchaud90}. In the strongly heterogeneous regime $0<\alpha<1$, the response to an external force $F$ crosses over from a linear regime, $\langle x(t)\rangle \propto F$, to a nonlinear regime, $\langle x(t)\rangle \propto F^{\alpha}$. The crossover between these behaviors is itself observation-time dependent: the threshold force scales as 
\begin{equation}
F_c(t)\propto t^{-\alpha/(1+\alpha)}, 
\end{equation}
indicating that the apparent boundary between linear and nonlinear response shifts to smaller forces as the observation time increases, demonstrating an observation-time-induced crossover in the mean displacement.

This observation-time dependence can also be understood within the present framework. In the driven QTM with $0<\alpha<1$, the mean displacement scales as
\begin{align}
\langle x(t)\rangle \sim \varepsilon \langle N_t\rangle ,
\end{align}
where $\varepsilon$ represents the bias strength. In the strongly heterogeneous regime the renewal number follows $\langle N_t\rangle \sim t^{\alpha}$, whereas the bias modifies the effective trapping statistics and produces a nonlinear contribution scaling as $\langle x(t)\rangle \sim (\varepsilon t)^{\alpha}$. Balancing the linear and nonlinear contributions yields the crossover bias
\begin{align}
\varepsilon_c(t)\propto t^{-\alpha/(1+\alpha)},
\end{align}
which coincides with the scaling reported in Ref.~\cite{bouchaud90}. Interestingly, this observation-time-dependent crossover in the mean displacement originates from the same mechanism as the variance crossover discussed in the present work, namely the competition between the observation time and a bias-dependent relaxation timescale of the transport response.

Related finite-time crossovers also arise in systems where the asymptotic behavior is controlled not by disorder-induced relaxation, but by the delayed emergence of nonlinear effects. One example is stochastic interface growth in the Kardar--Parisi--Zhang universality class \cite{PhysRevLett.56.889,HALPINHEALY1995215}. There, nonlinear effects controlled by a parameter $\lambda$ become visible only beyond a characteristic timescale. Finite-time analyses show that the effective crossover scale depends on the observation time, yielding in one dimension $\lambda_c(t)\propto t^{-1/4}$ \cite{PhysRevA.45.7156}. 
These examples suggest that observation-time-induced crossovers may arise
whenever the characteristic relaxation timescale controlling the asymptotic
dynamics depends on an external control parameter.

Similar observation-window-dependent crossovers have also been reported experimentally. A representative example is the protein dynamical transition in hydrated proteins, which is usually identified as a sharp increase in the measured mean-square displacement over a certain temperature range \cite{doster1989dynamical,Doster1990,Tournier2003,Becker2004,roh2006influence,Becker2003,Vural2013,Doster2008Thedynamical,doster2010protein,Magazu2011,Ngai2017}. The apparent transition temperature is known to depend on the instrumental observation window, or equivalently the resolution time \cite{Magazu2011, Ngai2017, Becker2004, Doster2008Thedynamical}. This suggests that the observed crossover is governed not only by temperature itself but also by the competition between a temperature-dependent relaxation process and the available observation window. Although the microscopic origin of the relevant relaxation remains under debate, it has often been discussed in connection with hydration-water-related dynamics and, more generally, protein--water coupled motions \cite{schiro2015translational, Magazu2011}.

It is also worth noting that slow relaxation can arise from different types of heterogeneity. 
In systems with temporally fluctuating diffusivity, as discussed in the framework of diffusing diffusivity 
\cite{Chubynsky2014,Chechkin2017,kimura2023non,Akimoto2026}, the effective diffusion coefficient has been shown to exhibit an 
observation-time-induced crossover with respect to temperature \cite{Shirataki2026}. In this case, slow relaxation can arise from temporal heterogeneity.

A relevant experimental realization of CTRW dynamics is provided by charge transport in disordered materials, as originally demonstrated by Scher and Montroll \cite{Scher1975}, where broad waiting-time distributions arise from trapping processes. More recent realizations include tracer particles in heterogeneous soft-matter systems, where intermittent trapping and release lead to similar anomalous transport behavior \cite{Wong2004,Weigel2011}. In such systems, a weak external force can be applied, for instance in active microrheology experiments, and the displacement variance can be measured over finite observation times.
Although our analysis is presented in one dimension, the underlying mechanism depends only on the renewal statistics of the number of steps. In higher-dimensional systems, the motion along the direction of the applied bias can be projected onto one dimension, leading to the same effective description. 

Related behavior may also be explored in structured media such as comb-like geometries, where motion along a backbone is intermittently interrupted by excursions into side branches, effectively generating broad waiting-time distributions. In such systems, applying a weak bias along the backbone and measuring the displacement variance over finite observation times would provide another route to test the predicted crossover.

It is also instructive to consider how the present results are modified in the presence of aging, particularly for $0<\alpha<1$, where the mean waiting time diverges. 
If an aging time $t_a$ is introduced, the relevant renewal number becomes $N(t_a+t)-N(t_a)$, whose statistics depend explicitly on both the observation time $t$ and the age $t_a$ \cite{Schulz2013,*Schulz2014,Akimoto2023}. 
In this case, the competition underlying the observation-time-induced crossover is expected to involve both times, leading to a generalized threshold $\varepsilon_c(t,t_a)$ that depends on the ratio $t/t_a$. 
Physically, aging tends to suppress the response, as older systems are more likely to remain trapped in long waiting states, making the bias-induced contribution harder to detect within a finite observation window. 
A systematic analysis of such aging-induced modifications of the crossover is an interesting direction for future work.

To place the present results in a broader context, we relate them to continuous stochastic descriptions. In particular, Langevin dynamics in random potentials under external driving can, under suitable coarse-graining, be mapped onto a trap-model description, where escape from each potential well is governed by Arrhenius-type dynamics. In this sense, the driven QTM can be viewed as an effective description of activated transport in disordered energy landscapes.
In such systems, the presence of a parameter-dependent relaxation timescale can delay the emergence of nonequilibrium response, and the apparent crossover depends on the observation time.
Similar phenomenology has been reported in Langevin-type systems in periodic potentials under time-dependent or nonequilibrium driving \cite{Spiechowicz2021,spiechowicz2016transient,spiechowicz2019squid}. Although these systems do not involve quenched disorder, they exhibit slow relaxation and finite-time effects that lead to observation-window-dependent crossover behavior.

\section{Conclusion}

We studied how a weak constant bias becomes detectable in anomalous transport through nonequilibrium fluctuations, focusing on the displacement variance in the driven CTRW and the driven QTM with power-law waiting-time tails.
Using an exact identity that links the displacement variance to the renewal statistics, we showed that for $\alpha<2$ the normalized transport response
$R(t,\varepsilon)=\mathrm{Var}[x(t)]_{\varepsilon>0}/\mathrm{Var}[x(t)]_{\varepsilon=0}$
exhibits a clear observation-time-induced crossover:
for sufficiently weak bias, the variance remains effectively indistinguishable from the unbiased behavior, whereas above a threshold bias $\varepsilon_c(t)$ it crosses over to a bias-induced nonequilibrium scaling.
The threshold decreases with the observation time, implying that longer observations reveal weaker driving through fluctuations.

We derived the observation-time scaling of the threshold bias $\varepsilon_c(t)$ in each $\alpha$ regime and compared the behaviors of the driven CTRW and the driven QTM.
For the driven CTRW, the crossover arises from the competition between the unbiased contribution and the bias-amplified renewal-number fluctuations, yielding $\varepsilon_c(t)\propto t^{-\nu}$ with $\nu=\alpha/2$ for $0<\alpha<1$ and $\nu=(2-\alpha)/2$ for $1<\alpha<2$.
For the driven QTM, quenched disorder makes the renewal statistics explicitly bias dependent, leading to distinct exponents and, notably, a larger $\nu$ than in the CTRW for the same $\alpha$.
This implies that, at fixed $(t,\alpha)$, the QTM exhibits a detectable response at smaller bias than the CTRW.

From a mechanistic perspective, the crossover can be interpreted as a competition between the observation time and a bias-dependent relaxation time of the transport response, $t_{\rm resp}(\varepsilon)$.
The emergence of an observation-time-induced crossover therefore requires that the relaxation toward the bias-induced scaling is controlled by an external parameter.
Our results provide a simple and quantitative framework to diagnose such protocol-dependent detectability of weak driving in heterogeneous transport.
An important future direction is to go beyond the present mean-field treatment of biased QTM renewal statistics and to extend the analysis to higher dimensions and more complex disorder landscapes, where the interplay of bias, revisits, and heterogeneity can further enrich the finite-time response.

\section*{Acknowledgement}
T.A. was supported by JSPS Grant-in-Aid for Scientific Research (No.~C 21K033920).



\bibliography{akimoto}

\end{document}